\documentclass[12pt]{iopart}

\usepackage[colorlinks=true,linkcolor=blue,urlcolor=blue,citecolor=blue,pdfusetitle]{hyperref}
\usepackage{color}
\usepackage[utf8]{inputenc}
\usepackage[usenames,dvipsnames]{xcolor}
\usepackage{amssymb}
\usepackage{graphicx}
\graphicspath{{graphics/}}
\usepackage{epsfig}
\usepackage{dcolumn}
\usepackage{bm}
\usepackage{mathrsfs}
\usepackage{multirow}
\usepackage[all]{xy}
\usepackage{pbox}
\usepackage{verbatim}
\usepackage{braket}
\usepackage{bbold}
\usepackage{xr}
\usepackage{float}
\usepackage{stackrel}

\newcommand{\eqref}[1]{(\ref{#1})}

\begin{document}
\raggedbottom

\title{Quantum thermodynamics of boundary time-crystals
}

\author{Federico Carollo$^1$, Igor Lesanovsky$^{1,2}$, Mauro Antezza$^{3,4}$, and Gabriele De Chiara$^{5,4,*}$}
\address{$^1$Institut f\"{u}r Theoretische Physik,  Universit\"{a}t T\"{u}bingen, Auf der Morgenstelle 14, 72076 T\"{u}bingen, Germany}
%\author{Igor Lesanovsky}
\address{$^2$ School of Physics and Astronomy and Centre for the Mathematics and Theoretical Physics of Quantum Non-Equilibrium Systems, The University of Nottingham, Nottingham, NG7 2RD, United Kingdom}

%\author{Mauro Antezza}
\address{$^3$ Laboratoire Charles Coulomb (L2C) UMR 5221 CNRS-Universit\'e de Montpellier, F- 34095 Montpellier, France}
\address{$^4$ Institut Universitaire de France, 1 rue Descartes, F-75231 Paris Cedex 05, France}
%\author{Gabriele De Chiara}
\address{$^5$ Centre for Quantum Materials and Technology, School of Mathematics and Physics, Queen’s University Belfast, Belfast BT7 1NN, United Kingdom}
\address{$^*$ Corresponding author: g.dechiara@qub.ac.uk}

\date{\today}

\begin{abstract}
    Time-translation symmetry breaking is a mechanism for the emergence of non-stationary many-body phases, so-called time-crystals, in Markovian open quantum systems. Dynamical aspects of time-crystals have been extensively explored over the recent years. However, much less is known about their thermodynamic properties, also due to the intrinsic nonequilibrium nature of these phases. Here, we consider the paradigmatic boundary time-crystal system, in a finite-temperature environment, and demonstrate the persistence of the time-crystalline phase at any temperature. Furthermore, we analyze thermodynamic aspects of the model investigating, in particular, heat currents, power exchange and irreversible entropy production. 
    Our work sheds light on  the thermodynamic cost of sustaining nonequilibrium time-crystalline phases and  provides a framework for characterizing time-crystals as possible resources for, e.g., quantum sensing. Our results may be verified in experiments, for example with trapped ions or superconducting circuits, since we connect thermodynamic quantities with mean value and covariance of collective (magnetization) operators.
\end{abstract}

\maketitle

\section{Introduction}
Nonequilibrium quantum systems can host a wide range of collective phenomena, including self-organized criticality \cite{kosmas2018,helmrich2020}, synchronization \cite{giorgi2019,karpat2021} and phase transitions even in low dimensions \cite{hinrichsen2000,kshetrimayum2017,carollo2019,kalthoff2022,chertkov2022}. By combining periodic driving, disorder, and strong interactions, unique ``time-crystalline" phases  \cite{WilczekPRL2012} emerge, wherein the discrete time-translation symmetry of the driving is broken. This manifests in temporal correlations at multiples of the driving period~\cite{sacha2015,khemani2016,else2016,SachaRPP2018,khemani2019brief,Else2020}. These phases, known as discrete time-crystals, have been recently observed in several experiments, e.g., with disordered spin systems~\cite{ChoiNature2017, randall2021many}, trapped ions~\cite{zhang2017observation}, and superconducting circuits~\cite{FreySA2022}.

In Markovian settings, time-crystals arise as a consequence of the breaking of the continuous time-translation symmetry of the (time-independent) dynamical generator. These instances manifest through  asymptotic self-sustained oscillations associated with the approach of the state of the system to a limit-cycle dynamics. Since time-crystals are not possible in equilibrium \cite{bruno2013,watanabe2015}, i.e., for systems in their ground states or thermalizing at a finite temperature, research work on continuous time-crystals is mostly focused on nonequilibrium (open quantum) settings. A paradigmatic model displaying a time-crystalline phase is the {\it boundary time-crystal} (BTC) \cite{iemini2018}. It consists of an ensemble of spins subject to an external field and undergoing collective dissipative dynamics. When the external field is sufficiently strong, the system state can enter persistent oscillations. The emergence of such a limit-cycle dynamics can be traced back to the presence, in the thermodynamic limit, of a large decoherence-free subspace stabilized by the closure of the gap of the dynamical generator \cite{iemini2018}. This phenomenology can be formalized via the concept of dynamical symmetries \cite{buca2019,sanchezmunoz2019,buca2022}. 

Despite these profound insights on their emergent dynamics, much less is known about thermodynamic properties of time-crystalline phases (see related issues for nonequilibrium engines \cite{carollo2020a,carollo2020,paulino2022}). 
Understanding and controlling heat currents, power exchanges and irreversible entropy production in these systems is, however, both of fundamental interest and of practical relevance, for instance for exploring efficiency measures and thermodynamic costs of possible applications of time-crystals as resources for quantum sensing~\cite{montenegro2023,pavlov2023,iemini2023}. 
The main challenge for such a quantum thermodynamic analysis lies in the fact that, given that these systems are genuinely out of equilibrium, they are not described by thermal open quantum dynamics \cite{breuer2002}. They are  instead governed by so-called {\it local} quantum master equations, implementing genuine nonequilibrium time evolutions \cite{HewgillPRR2021,DeChiaraPRR2022}, which may lead to thermodynamic inconsistencies~\cite{levy2014, stockburger2017} if a suitable microscopic analysis is not performed~\cite{barra2015,StrasbergPRX2017,DeChiaraNJP2018}.

Here, we develop, to the best of our knowledge for the first time, a coherent thermodynamic description of an open quantum time-crystal. We consider a finite-temperature BTC, based on a local Markovian quantum master equation, and derive the exact dynamics of the average magnetization and of the associated quantum fluctuations. As we demonstrate, BTCs are robust to thermal effects and can be observed also in the infinite-temperature limit. In particular, the average magnetization is independent of the external temperature which merely increases its fluctuations. Exploiting a collision-model description for the system-environment interaction \cite{CampbellEPL2021,CiccarelloPR2022,cusumano2022,li2023} (see sketch in Fig.~\ref{fig:setup}), heat currents, work power and entropy production for the BTC can be consistently defined. These  quantities 
exhibit nonanalytic behavior across the nonequilibrium time-crystal phase transition, with maximal absorbed power and critical dynamics occurring at the phase-transition point. 
Importantly, all relevant quantities (including the von Neumann entropy) can be expressed in terms of quantum observables, which makes it possible to verify our results in experiments. 
\\

\section{The model}
We consider the BTC introduced in Ref.~\cite{iemini2018}, later generalized in Refs.~\cite{sanchezmunoz2019,piccitto2021,prazeres2021,passarelli2022,passarelli2023}, in the presence of a finite-temperature environment. The system consists of an ensemble of $N$ spin-$1/2$ particles which is described by the collective spin operators $V_\alpha=\sum_{k=1}^N\sigma_\alpha^{(k)}/\sqrt{2}$, with $\sigma_\alpha^{(k)}$ being the $\alpha$th Pauli matrix for spin $k$. These operators satisfy the commutation relation $[V_\alpha,V_\beta]=i\sqrt{2}\epsilon_{\alpha\beta\gamma}V_\gamma$, where $\epsilon$ is the Levi-Civita tensor. 

The time evolution of the state of the system, $\rho(t)$, is implemented by the Markovian quantum master equation $\dot{\rho}(t)=\mathcal{L}[\rho(t)]$, with Lindblad generator \cite{lindblad1976,gorini1976,breuer2002} 
\begin{eqnarray}
\mathcal{L}[\rho]:&=&-i[H,\rho]+\frac{\Gamma(n_\beta+1)}{N}\left(V_-\rho V_+-\frac{1}{2}\left\{\rho,V_+V_-\right\}\right)
\nonumber\\
&+&\frac{\Gamma n_\beta}{N}\left(V_+\rho V_--\frac{1}{2}\left\{\rho ,V_-V_+\right\}\right)\, .
    \label{Lindblad}
\end{eqnarray}
Here, $n_\beta=(e^{\beta\omega}-1)^{-1}$ is the thermal occupation of the environmental particles, $\omega$ their 
energy scale and $\beta$ is the bath's inverse temperature.
The Hamiltonian of the system is $H=(\Omega/\sqrt{2}) V_x$, with $\Omega$ being a transverse field. The operators $V_\pm=V_x\pm iV_y$  are  collective ladder operators and $\Gamma$ represents an overall decay rate. The  factor $1/N$ in the dissipative terms of the above equation ensures a well-defined thermodynamic limit \cite{iemini2018,benatti2016,benatti2018}. The original BTC of Ref.~\cite{iemini2018} is retrieved in the zero-temperature ($\beta\to\infty$) limit. Here, a nonequilibrium phase transition emerges from a stationary to a time-crystal phase, in which the system features sustained oscillations  \cite{iemini2018,CarolloPRA2022}. In what follows, we study the system for finite temperatures and derive its quantum thermodynamics.\\

\subsection{Mean-field description and time-crystal phase}   
We introduce the average magnetization operators, defined as $m_\alpha^N=V_\alpha/N$, providing a useful set of order parameters for nonequilibrium phase transitions. Whenever considering states with sufficiently low correlations, often referred to as clustering states, these operators converge  to multiples of the identity, $\lim_{N\to\infty} m_\alpha^N=\lim_{N\to\infty}\langle m_\alpha^N\rangle=m_\alpha$, where $m_\alpha$ is the limiting expectation value of $m_\alpha^N$ in the reference state  \cite{landford1969,bratteli1982,thirring2013,strocchi2021}. This convergence is nothing but a law of large numbers and implies that, whenever considering expectation values of products of operators on a clustering state,  average magnetizations can be considered as multiples of the identity in the thermodynamic limit (see 
a more rigorous discussion, for instance, in Ref,~\cite{benatti2018}). Under the dynamics implemented by the Lindblad generator in Eq.~\eqref{Lindblad}, average magnetization operators remain proportional to the identity at all times and evolve according to the mean-field equations \cite{iemini2018,benatti2018,carollo2021} (see details in \ref{App1})
\begin{eqnarray}
\dot{m}_x(t)&=&\sqrt 2 \Gamma m_z(t) m_x(t)\, ,
\\
\dot{m}_y(t)&=&m_z(t)[\sqrt 2 \Gamma m_y(t)-\Omega]\, ,
\\
\dot{m}_z(t) &=& \Omega m_y(t) -\sqrt 2 \Gamma[m_x^2(t)+m_y^2(t)] \, .
\label{mean-field-eq}
\end{eqnarray}
These equations feature two conserved quantities. The first, $m^2=m_x^2(t)+m_y^2(t)+m_z^2(t)$, emerges in the limit $N\to\infty$ due to conservation of the total spin operator and we fix it to $m^2=1/2$. The second conserved quantity is $c=m_x(t)/[m_y(t)-\Omega/(\sqrt{2}\Gamma)]$  and is also only present in the thermodynamic limit. Looking at Eq.~\eqref{mean-field-eq}, we see that the dynamics of the average magnetization operators do not depend on the temperature. This implies that, whenever $\Omega>\Gamma$, irrespectively of the temperature, the state of the system enters the time-crystal phase, as witnessed by average operators approaching a limit-cycle dynamics \cite{iemini2018,CarolloPRA2022}. The robustness of the time-crystal phase against finite temperatures is our first main result.

\subsection{Quantum fluctuation operators}  
Going beyond average magnetization operators, it is possible to provide an exact description for the so-called quantum fluctuation operators \cite{goderis1989,goderis1990,narnhofer2002,benatti2016,benatti2018}. In analogy with classical central limit theorems, we can indeed introduce quantum fluctuations as
\begin{equation}
    F_\alpha^N=\frac{1}{\sqrt{N}}\left[V_\alpha -\langle V_\alpha\rangle\right]\, .
    \label{qfo}
\end{equation}
These operators quantify the deviation of $V_\alpha$ from the average value calculated from the reference state whose quantum fluctuations are of interest. These operators are called quantum fluctuations as they are constructed from the microscopic spin operators. However they account for both quantum and thermal fluctuations.
Analogously to what happens in  classical settings, fluctuations converge, in the thermodynamic limit, to operators $F_\alpha$ equipped with a Gaussian  quantum state for which they assume a zero average value and possess a variance given by $\langle F_\alpha^2\rangle =\lim_{N\to\infty}\langle (F_\alpha^N)^2\rangle$.
For quantum systems, the fluctuations $F_\alpha$   can in addition retain a quantum character in the thermodynamic limit and generically behave as bosonic operators \cite{goderis1989,goderis1990}. Their commutation relations $[F_\alpha,F_\beta]=is_{\alpha\beta}$, where we have introduced the (degenerate) $3\times3$ symplectic matrix $s_{\alpha\beta}=\sqrt{2}\epsilon_{\alpha\beta\gamma}m_\gamma$, depend on the values of the average magnetizations. In the large-$N$ limit, the state of the quantum fluctuations is fully characterized by the $3\times3$ covariance matrix  
    $G_{\alpha\beta}=\frac{1}{2}\langle \left\{F_\alpha,F_\beta\right\}\rangle$, which essentially contains the susceptibility of the order parameters.
Under the time evolution in Eq.~\eqref{Lindblad}, the state of the quantum fluctuations remains Gaussian  and the covariance-matrix dynamics can be computed explicitly \cite{benatti2016,benatti2018,CarolloPRA2022,nie2023} (see details in  \ref{App1}). 
The combined analysis of average magnetizations and quantum fluctuations allows us to determine the behavior of all relevant thermodynamic quantities, as we demonstrate in  \ref{App1}. 
\\

\section{Thermodynamic description of the BTC}
\subsection{Collision models}
The dissipative term in Eq.~\eqref{Lindblad} does not lead to transitions between eigenstates of the Hamiltonian $H$ \cite{breuer2002}. Indeed, the ladder operators $V_\pm$ map between eigenstates of $V_z$ while the Hamiltonian of the system is proportional to $V_x$. These master equations are broadly referred to as \emph{local} master equations and lead to inconsistencies whenever considering a thermodynamic interpretation based on a weak system-bath coupling \cite{barra2015,levy2014,stockburger2017,DeChiaraNJP2018}. To provide a consistent thermodynamic framework for the BTC, we thus consider a different realization of the system-environment interaction \cite{barra2015,DeChiaraNJP2018,StrasbergPRX2017}, also known as collision-model dynamics  \cite{CampbellEPL2021, CiccarelloPR2022}. 

\begin{figure}[htbp]
\begin{center}
\includegraphics[width=0.7\columnwidth]{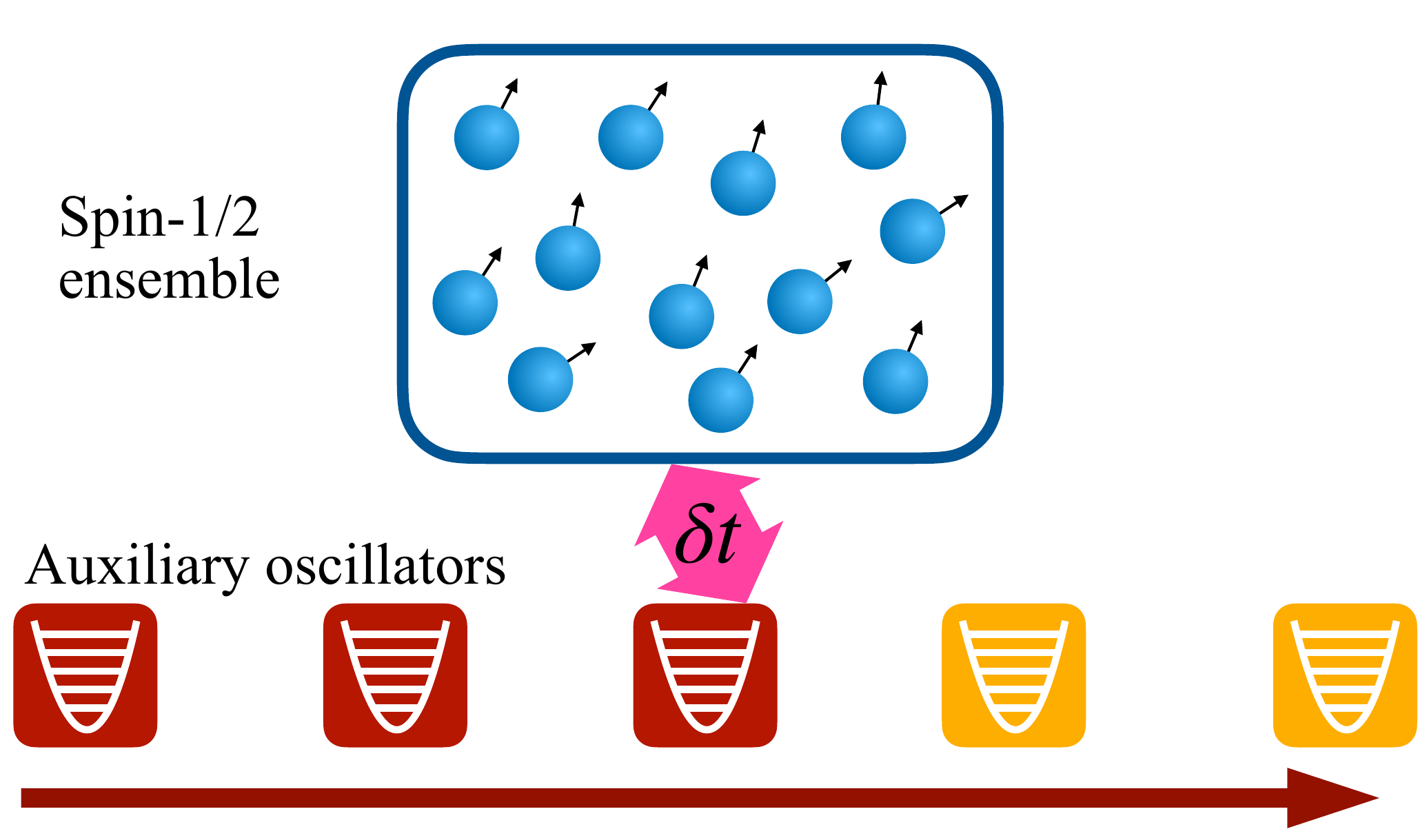}
\caption{{\bf Collision-model setup.}  The system of interest consists of an ensemble of spin-1/2 particles, which collides with a sequence of auxiliary quantum harmonic oscillators. The latter represent the environmental particles and are initially prepared in an equilibrium thermal state. The system-environment interaction lasts for an infinitesimal time $\delta t$ and, after the collision, the auxiliary oscillator is discarded. In the limit $\delta t\to 0$, the reduced dynamics of the spin-$1/2$ ensemble is described by the quantum master equation in Eq.~\eqref{Lindblad}.}
\label{fig:setup}
\end{center}
\end{figure}

In the setup that we consider, which is illustrated in Fig.~\ref{fig:setup}, the environment (E) is composed by a stream of auxiliary units, here assumed to be quantum harmonic oscillators, which interact sequentially with the spin system (S) for a short time $\delta t$ \cite{CiccarelloPR2022,CampbellEPL2021}. We denote with $a_k,a^\dagger_k$ the annihilation and creation operators for the $k$th environmental particle, fulfilling canonical commutation relations. The composite system-environment Hamiltonian is $H_{\rm tot}(t)=H+\sum_{k=1}^\infty H_k(t)$, where $H_k(t)=\omega a_k^\dagger a_k+H_k^{\rm int}(t)$ and with
\begin{equation}
H_k^{\rm int}(t)=\sqrt{\frac{\Gamma}{N\delta t}} g[t-(k-1)\delta t]\left(a_k^\dagger V_-+a_kV_+\right)
    \label{H_int}
\end{equation}
describing the interaction between the system and the $k$th environmental particle. The function $g(t)$, is defined as $g(t)=1$ for $0<t<\delta t$ and 0 otherwise. 
The environment is initialized with all particles in the thermal state $\rho_k^{\rm E}\propto  e^{-\beta \omega a_k^\dagger a_k}$.  

%%%%%%%%%%%%%%%%%%%%%%%%%%%%%

%%%%%%%%%%%%%%%%%%%%%%%%%%%%%

The reduced state of the system after the interaction with the $k$th environmental particle is obtained through the iterative discrete-time equation  
\begin{equation}
\rho(t+\delta t)={\rm Tr}_k \left [ U_k \rho(t) \otimes \rho_k^{\rm E} U_k^\dagger \right]\, , 
\end{equation}
where $t= (k-1)\delta t$, $U_k=e^{-i\int_t^{t+\delta t}{\rm d }s \, \left[H+H_k(s)\right]}$ and ${\rm Tr}_k$ denotes the trace over the $k$th environmental particle.  
Following Ref.~\cite{DeChiaraNJP2018},  expanding the unitary operators $U_k$ up to first order in $\delta t$, taking the continuous limit $\delta t\to 0$ and calculating $\dot{\rho}(t)=\lim_{\delta t\to 0} [\rho(t+\delta t)-\rho(t)]/\delta t$, we recover the evolution generated by Eq.~\eqref{Lindblad}.\\

\subsection{Quantum thermodynamics}
Building on this microscopic model, [cf.~sketch in Fig.~\ref{fig:setup}], we now derive all thermodynamic quantities for the BTC and analyze their behavior across the nonequilibrium time-crystal phase transition, which constitute our second main result.

The heat exchanged between the system and the bath in the time-interaval $t\to t+\delta t$, with $t=(k-1)\delta t$, is obtained as (minus) the energy absorbed by the $k$th environmental particle during the interaction, or collision, with the system 
\begin{eqnarray}
\dot{Q}(t) &=& - \lim_{\delta t\to 0} \frac{1}{\delta t} {\rm Tr}_k \left\{ \omega \, a_k^\dagger a_k \Delta\rho_k^{\rm SE} \right\}=
\nonumber
\\
&=& \frac{\omega\Gamma}{N} \left [ n_\beta \langle V_-V_+ \rangle - (n_\beta+1) \langle V_+V_- \rangle \right ],
\label{eq:heat}
\end{eqnarray}
where $\Delta\rho_k^{\rm SE}=U_k \rho(t) \otimes \rho_k^{\rm E} U_k^{\dagger}-\rho(t) \otimes \rho_k^{\rm E}$. We use the convention that heat flowing into the system is positive. 
The above heat current vanishes only if the spin ensemble thermalizes, i.e., when $
\langle V_+V_- \rangle/\langle V_-V_+ \rangle=n_\beta/(n_\beta+1)$.

The internal energy of the system, $U(t)=\langle H\rangle$, obeys the relation 
\begin{equation}
\dot{U}(t) =\frac{\Omega \Gamma}{N} \left[\frac{\langle V_xV_z+V_z V_x \rangle }{2}-\frac{2n_\beta+1}{\sqrt 2}\langle V_x\rangle    \right ].
\end{equation}
Since the interaction in Eq.~\eqref{H_int} is time-dependent, there is also a  (work-)power input \cite{DeChiaraNJP2018}, given by 
\begin{equation}
\dot{W}(t)= \lim_{\delta t\to 0}  \frac{1}{\delta t} {\rm Tr}_k \left\{ (H+ \omega a_k^\dagger a_k )\Delta\rho^{\rm SE}_k\right\}=\dot{U}(t)-\dot{Q}(t)\, .
\label{eq:1stlaw}
\end{equation}
The  last expression is the first law of thermodynamics and we consider work to be positive when performed on the system.

\begin{figure}[t]
\centering
\includegraphics[width=0.8\columnwidth]{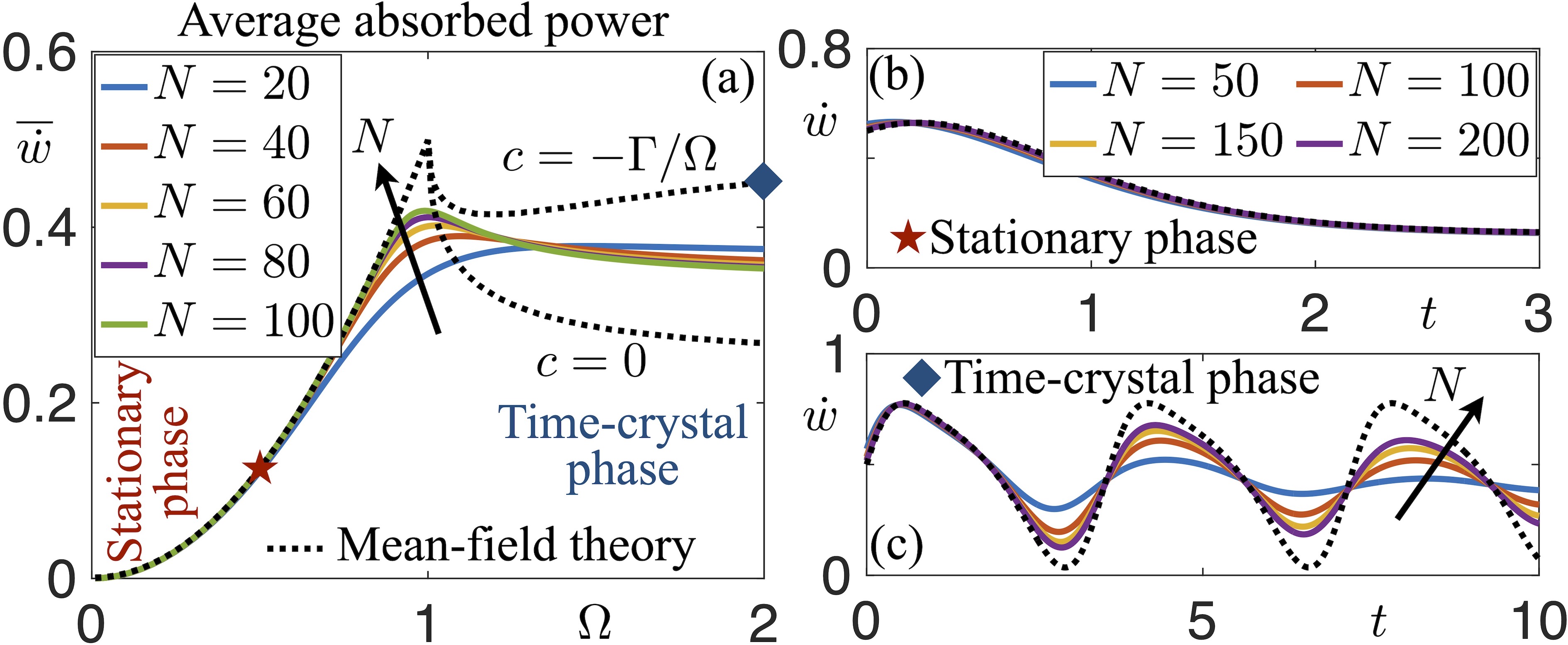}
\caption{{\bf Stationary power absorption.} (a) The dotted lines provide the mean-field prediction obtained by time-integrating the first law over a long time window. The latter depends on the value of the constant of motion $c$, in the time-crystalline phase. Solid lines are finite-$N$ results obtained from the stationary state of the master equation. These curves appear to converge to the mean-field prediction for $c=0$ (see also discussion in Ref.~\cite{cabot2022}) (b) In the stationary phase ($\Omega/\Gamma=0.5$), the absorbed power per particle $\dot{w}(t)$ converges to a stationary value. (c) In the time-crystalline phase ($\Omega/\Gamma=2$), $\dot{w}(t)$ shows persistent oscillations. Finite-$N$ results converge to the prediction upon increasing $N$.  Power is in units of $\omega \Gamma$, $\Omega$ is in units of $\Gamma$ and time is in units of $\Gamma^{-1}$. Numerical data are obtained for $n_\beta=1$.  The initial state is the ground state of $H$, apart from the case $c=0$ in which it is the ``ground state" of $V_z$. }
\label{Fig1}
\end{figure}

To explore the behavior of these quantities in the thermodynamic limit, we introduce the internal energy per particle, $u(t)=U(t)/N$. The latter 
converges to a stationary value for any finite $N$ \cite{iemini2018}. In the large-time limit, the first law thus reads $\dot{w}=-\dot{q}$, where $w=W/N$ and $q=Q/N$. In Fig.~\ref{Fig1}(a), we show the stationary value of $\dot{w}$ [see also Fig.~\ref{Fig1}(b)] as a function of the transverse field $\Omega$. Such quantity develops a cusp for increasing values of $N$. 
In the thermodynamic limit and in the time-crystal phase, $\dot{w}(t)$ features instead persistent oscillations [cf.~Fig.~\ref{Fig1}(c)]. As such, rather than considering its instantaneous behavior, we consider its time-integrated average,
$$
\overline{\dot{w}}(t)=\frac{1}{t}\int_0^{t}{\rm d}s \, \dot{w}(s)\, ,
$$
which essentially provides, in the large-time limit, the average power absorbed over a single period of the oscillations. This quantity converges to the instantaneous $\dot{w}(t)$, for $t\to \infty$, in the stationary regime. We also have that $\lim_{t\to\infty}\overline{\dot{w}}(t)=-\lim_{t\to\infty}\overline{\dot{q}}(t)$, since $\lim_{t\to\infty}\overline{\dot{u}}(t)=0$, given that $u(t)$ is an oscillatory (bounded) function. Looking at Eq.~\eqref{eq:heat} and considering that $\langle V_{\mu}V_{\nu}\rangle/N^2\to m_\mu m_\nu$ (see \ref{App1}), we further see that, in the thermodynamic limit, heat currents and power per spin are solely determined (at leading order) by the mean-field behavior and do not depend on the temperature. As shown in Fig.~\ref{Fig1}(a), in the thermodynamic limit, the quantity $\overline{\dot{w}}$ shows a nonanaliticity at the nonequilibrium time-crystal phase transition, where the absorbed power attains its maximal value. The latter statement follows from the fact that $\dot{q}=-\omega\Gamma (m_x^2+m_y^2)$, for $N\to\infty$, that $m_x^2+m_y^2\le 1/2$ due to the conservation law, and that $m_z$ approaches zero at the critical point \cite{iemini2018,CarolloPRA2022}.

%%%%%%%%%%%%%%%%%%%%%%%%%%%%%%%%%%%%%%%%%%%%%%%%%%
%%%%%%%%%%%%%%%%%%%%%%%%%%%%%%%%%%%%%%%%%%%%%%%%%%

We now consider the irreversible entropy production in the system. To this end, we define the post-collision state $\rho_k^{\prime {\rm SE}}:=U_k \rho(t) \otimes \rho_{k}^{\rm E} U_k^{\dagger}$, $\rho^{\prime {\rm S}}:={\rm Tr}_k \rho_k^{\prime {\rm SE}}=\rho(t+\delta t)$ and $\rho_k^{\prime {\rm E}}:={\rm Tr} \rho_k^{\prime {\rm SE}}$. The entropy production can be defined as (see Ref.~\cite{LandiRMP2021, StrasbergPRXQuantum21})
%\begin{equation}
$\Sigma = I_k({\rm S}:{\rm E})+S(\rho^{\prime {\rm E}}_k || \rho^{ \rm E}_k) \ge 0$. 
%\end{equation}
Here, we have defined the mutual information between the system and the environmental unit after the $k$th collision (being zero before the collision)
    $I_k({\rm S}:{\rm E})=S(\rho^{\prime {\rm S}})+S(\rho_k^{\prime {\rm E}})-S(\rho_k^{\prime {\rm SE}})$,
and the relative entropy 
$S(\sigma || \rho) = {\rm Tr} \left[\sigma \ln \sigma - \sigma\ln \rho \right]$.
The non-negativity of the entropy production is guaranteed by the positivity of the mutual information and of the relative entropy. 
Putting all together we obtain:
$\Sigma=\Delta S - \Phi$
where $\Delta S=S(\rho^{\prime {\rm S}})-S(\rho^{\rm S})$, with $\rho^{\rm S}:=\rho(t)$, is the entropy variation of the system and $\Phi= {\rm Tr} \left[(\rho^{\prime {\rm E}}_k-\rho^{\rm E}_k)\ln\rho^{\rm E}_k \right]$ is the so-called entropy flux \cite{LandiRMP2021}. In the continuous-time limit, $\delta t\to 0$, we get:
$\dot{\Sigma}(t) =\dot{S}(t)-\dot{\Phi}(t)$.
Since $\rho^{\rm E}_k$ is thermal, we can further simplify the latter quantity as
\begin{equation}
\dot{\Phi}(t)=-\beta \lim_{\delta t \to 0}\frac {1}{\delta t} {\rm Tr}_k \left[(\rho^{\prime {\rm E}}_k-\rho^{ \rm E}_k)\omega a_k^\dagger a_k\right ]
=\beta\dot{Q}(t)\, ,
\end{equation}
which shows that the entropy flux is directly proportional to the heat current. 

Particular care is required to derive the behavior of the von Neumann entropy. At the mean-field level, the state of the system can be described by a product state over all spins, whose single-spin reduced density matrix is defined by the average magnetization operators. Due to the choice of the conservation law $m^2=1/2$, such a state is furthermore pure (we also remark here that the mean-field description is independent of the temperature).
This implies that the extensive (time-independent) contribution to the entropy, coming from the mean-field state, is vanishing. The (intensive) leading-order behavior must then be captured by the quantum fluctuations. It can be obtained, as in Gaussian bosonic system \cite{adesso2014}, by introducing the matrix $M=isG$ and finding its eigenvalues. Since $s$ is degenerate antisymmetric and since $G$ is symmetric, these are given by $0,\pm \lambda$, with $\lambda>0$. The latter two eigenvalues are related to the canonical (bosonic) mode which can be defined from the quantum fluctuations (see, e.g., also Refs.~\cite{benatti2018,CarolloPRA2022}). The entropy can then be calculated as 
$
S=\lambda_+\log\lambda_+-\lambda_-\log\lambda_-\, ,
$
with $\lambda_\pm=\lambda\pm1/2$ (see details in \ref{App1}).

\begin{figure*}[t]
\centering
\includegraphics[width=0.9\textwidth]{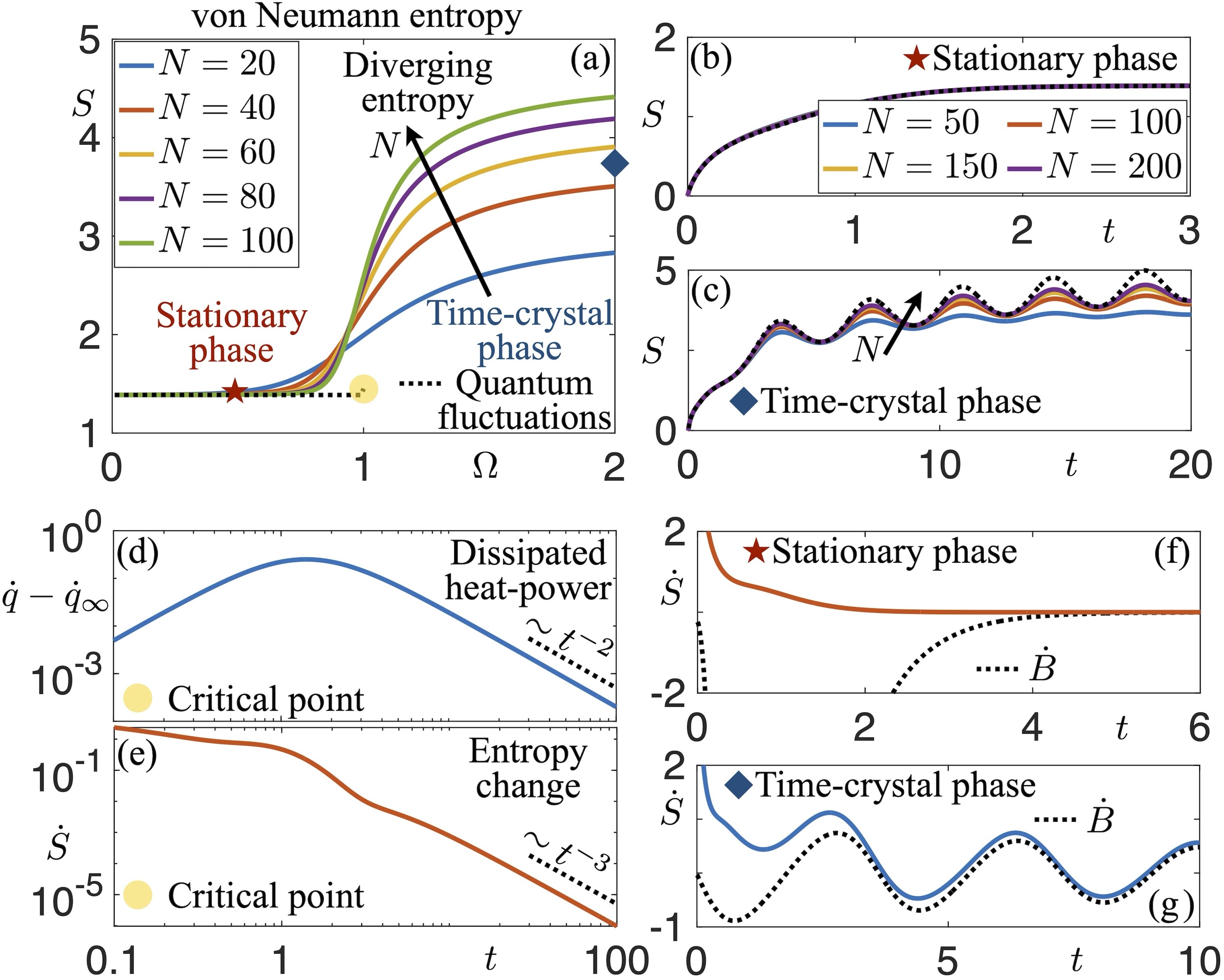}
\caption{{\bf Entropies and critical behavior. } (a) Stationary von Neumann entropy $S$. The dotted line is the prediction from quantum fluctuations, solely converging to a finite value in the stationary phase. Solid lines are finite-$N$ results obtained from the stationary state of the master equation. (b) Dynamics of the entropy in the stationary regime ($\Omega/\Gamma=0.5$) and comparison between prediction and finite-$N$ results. (c) Same as (b) in the time-crystal regime ($\Omega/\Gamma=2$). (d) At criticality, the dissipated heat-power approaches its stationary value $\dot{q}_\infty$ with a power-law behavior. (e) The von Neumann entropy change $\dot{S}$ also features a power-law decay, with a different exponent. (f-g) Comparison between $\dot{S}$ and $\dot{B}$ (see Eq.~\eqref{eq:Bdot}), for a system of $N=100$ spins, in the stationary regime, $\Omega/\Gamma=0.5$, (f) and for the time-crystal, $\Omega/\Gamma=2$, (g). Here, $n_\beta=1$ and heat-power is in units of $\omega\Gamma$. Time is in units of $\Gamma^{-1}$ and $\Omega$ in units of $\Gamma$. The initial state is the ground state of $H$.}
\label{Fig2}
\end{figure*}

As we show in Fig.~\ref{Fig2}(a-b), in the stationary regime, numerical results for the entropy $S$ converge to the prediction obtained via quantum fluctuations. The latter shows a constant entropy over the whole stationary phase, which is equal to the  one of a thermal Gaussian state, given by $S_\beta=(n_\beta+1)\log(n_\beta+1)-n_\beta\log n_\beta$. In this regime, the bosonic state of the quantum fluctuations is however not just a thermal state but is furthermore squeezed \cite{CarolloPRA2022}. On the other hand, in the time-crystal phase the entropy diverges logarithmically with time [cf.~Fig.~\ref{Fig2}(c)], in the thermodynamic limit. For finite systems, this manifests in a stationary entropy which increases  with $N$. At the critical point, heat currents, entropy flux, and the derivative of the von Neumann entropy show power-law behavior with time, as shown in Fig.~\ref{Fig2}(d-e), witnessing a slow decay of the heat current and of the absorbed power, as well as a critically slow approach towards the stationary state. The latter differs from the one in the stationary regime, as witnessed by a slightly different value of the entropy in Fig.~\ref{Fig2}(a).  

%%%%%%%%%%%%%%%%%%%%%%%%%%%%%%%%%%%%%%%%%%%%%%%%%%
%%%%%%%%%%%%%%%%%%%%%%%%%%%%%%%%%%%%%%%%%%%%%%%%%%

For completeness, we also show a bound on the entropy change $\dot{S}$, based on a result by Spohn~\cite{Spohn1978}. For any Lindblad generator $\mathcal L$, with steady state $\pi$, such that $\mathcal L [\pi]=0$, the following relation holds
\begin{equation}
-{\rm Tr}\left\{ \mathcal L[\rho] (\ln \rho -\ln \pi)\right\} \ge 0.
\end{equation}
Since the time derivative of the entropy can be cast in the form 
%\begin{equation}
$\dot S=-{\rm Tr}\left\{ \mathcal L[\rho(t)] \ln \rho(t) \right\}$,
the quantity
\begin{equation}
\dot B =-{\rm Tr}\left\{ \mathcal L[\rho(t)] \ln \pi \right\},
\label{eq:Bdot}
\end{equation}
provides a lower bound, 
$\dot{S}(t)\ge \dot{B}(t)$, as shown in Fig.~\ref{Fig2}(f-g). In \ref{App2}, we further analyze stochastic entropy production and (generalized) fluctuation theorems for the BTC, akin to Crooks theorem~\cite{CrooksPRA2008}, exploiting the framework of Ref.~\cite{KwonPRX2019}. \\

\section{Conclusions} We have derived a complete thermodynamic description for the paradigmatic boundary time-crystal. 
 The system is intrinsically an out-of-equilibrium system, which does not thermalise with its own environment and therefore does not possess a well-defined temperature. To develop a thermodynamic framework for such a system, we have thus exploited a collision-model interpretation of its interaction with the environment.  This allowed us to consistently explore  the behaviour of relevant thermodynamic quantities across the whole phase diagram as well as  near the phase transition.
The BTC is surprisingly robust to finite temperature and the heat currents and power absorption are temperature independent at leading order. The von Neumann entropy, as well as the entropy flux, depends instead on the temperature of the bath. In the stationary phase, the entropy is that of a thermal state, even though quantum fluctuations are additionally squeezed. In the time-crystal regime, the temperature determines the rate of growth of the entropy (as apparent from the dynamics of the covariance matrix reported in  \ref{App1}). 

Our prediction may be verified in experimental quantum platforms since  heat currents, power and entropy contributions are related to measurable quantities, such as average values and susceptibility parameters of magnetization operators. This observation,  as well as the ideas we have presented, holds true for generic thermal and nonthermal dynamics of collective systems, such as those discussed in Ref.~\cite{benatti2018} which further includes the case of all-to-all interactions in the Hamiltonian.  This fact enables more broadly connections between theoretical thermodynamic analysis on these models and realistic experiments.  In this regard, we note that the boundary time-crystal phase is robust against external perturbations \cite{iemini2018,carollo2020}. Moreover, this model and related ones have been observed in recent experiments (see, e.g., Refs.~\cite{kongkhambut2022,ferioli2023}), which indirectly confirms the robustness of time-crystal phases against unavoidable disorder effects which are present in realistic setups.

\ack
G.D.C. acknowledges the support by the UK EPSRC EP/S02994X/1, the Royal Society IEC\textbackslash R2\textbackslash 222003 and thanks the group Theory of Light-Matter and Quantum Phenomena of the Laboratoire Charles Coulomb for hospitality during his stay in Montpellier. G.D.C.~and M.A.~acknowledge support by the QuantUM Project at Universit\'e de Montpellier.  I.L.~and F.C.~acknowledge financial support from the Deutsche Forschungsgemeinschaft (DFG, German Research Foundation) through the Research Unit FOR 5413/1, Grant No.~465199066. This project has also received funding from the European Union’s Horizon Europe research and innovation program under Grant Agreement No.~101046968 (BRISQ) and the ESPRC UK, Grant No.~EP/V031201/1. F.C.~is indebted to the Baden-W\"urttemberg Stiftung for the financial support of this research project by the Eliteprogramme for Postdocs.

\appendix

\section*{Appendix}

\section{Description of the system in the thermodynamic limit}
\label{App1}
In this Section of the Supplemental Material, we provide details on the derivation of the mean-field equations, which are presented in the main text, and of the dynamics of quantum fluctuations. A rigorous proof of these results is given in Ref.~\cite{benatti2018}. We furthermore discuss how products of collective operators can be written in terms of these quantities in the thermodynamic limit and how the von Neumann entropy can be calculated from quantum fluctuations.

\subsection{Mean-field dynamics}

The starting point is the calculation of the action of the generator on the average magnetization operators. To this end, we introduce the dynamical generator for the evolution in the Heisenberg picture which, for our model, can also be cast in the following form 
\begin{equation}
\mathcal{L}^*[X]=i[H,X]+\sum_{\mu,\nu}\frac{A_{\mu\nu}}{2N}[[V_\mu,X],V_\nu]+i\sum_{\mu,\nu} \frac{B_{\mu\nu}}{2N} \left\{\left[V_\mu, X\right],V_\nu\right\}\, ,
\label{sm-mf:gen}
\end{equation}
with real matrices, $A=A^T$, $B=-B^T$, given by 
\begin{equation}
A=\Gamma (2n_\beta +1)
\left(
\begin{array}{ccc}
    1&0&0\\
    0&1&0\\
    0&0&0
\end{array}
\right )
% \begin{pmatrix}
%     1&0&0\\
%     0&1&0\\
%     0&0&0
% \end{pmatrix}
\, ,\qquad 
B={\Gamma}
%\begin{pmatrix}
\left(
\begin{array}{ccc}
0&-1&0\\
1&0&0\\
0&0&0
\end{array}
\right )
%\end{pmatrix}\, .
\end{equation}
Then, we calculate the action of such generator on an average magnetization operator $m_\alpha^N$, which gives 
\begin{equation}
\mathcal{L}^*[m_\alpha^N]=-\Omega\sum_{\mu}\epsilon_{x\alpha\mu} m_{\mu}^N-\sum_{\mu,\nu,\eta,\zeta}\frac{A_{\mu\nu}}{2N}\epsilon_{\mu\alpha\eta}\epsilon_{\eta\nu\zeta}m_\zeta^N-\sum_{\mu, \nu,\eta} \frac{B_{\mu\nu}}{\sqrt{2}}\epsilon_{\mu\alpha\eta}\left\{m_\eta^N,m_\nu^N\right\}\, .
\label{sm-mf:act_gen}
\end{equation}
This relation can now be used to obtain the time-evolution of the limiting operator $m_\alpha(t)=\lim_{N\to\infty}\langle m_{\alpha}^N\rangle_t$. Indeed, we have that 
$$
\dot{m}_\alpha(t)=\lim_{N\to\infty}\langle \mathcal{L}^*[m_\alpha^N]\rangle_t\, ,
$$
where the exchange of the limit and the derivation follows from the results of Ref.~\cite{benatti2018}. We can thus now calculate the expectation value with respect to the clustering state  $\langle \cdot\rangle_t$ (see again Ref.~\cite{benatti2018} or also Ref.~\cite{carollo2021}) of Eq.~\eqref{sm-mf:act_gen}. We note that the second term in Eq.~\eqref{sm-mf:act_gen} vanishes in norm, since $\|m_\alpha^N\|\le 1/\sqrt{2}$ and since there is a factor $1/N$ multiplying the elements of the matrix $A$. As such, this term does not contribute in the thermodynamic limit. For the third term, we note that since the expectation $\langle \cdot\rangle_t$ is with respect to a clustering state, average operators converge to multiples of the identity in the thermodynamic limit and, thus, $\lim_{N\to\infty}\langle m_\eta^N m_\nu^N\rangle_t=m_{\eta}(t) m_{\nu}(t)$ \cite{benatti2018,carollo2021}. With these observations, we find the mean-field equations 
$$
\dot{m}_\alpha(t)=-\Omega \sum_\mu \epsilon_{x\alpha \mu }m_{\mu}(t) -\sqrt{2}\sum_{{\mu\nu}}B_{\mu\nu}\epsilon_{\mu\alpha \eta} m_\eta(t) m_\nu(t)\, .
$$

\subsection{Dynamics of quantum fluctuations}
 Since we are interested in analysing the dynamical behavior of quantum fluctuation operators, we define them as  
$$
F_\mu^N= \frac{1}{\sqrt{N}}(V_\mu -\langle V_\mu\rangle_t)\, ,
$$
where $\langle \cdot \rangle_t$ denotes expectation with respect to the time-evolved quantum state. In this way, we also have that $\langle F_\mu^N\rangle_t=0$. We note that for the purpose of this work, we are only considering initial product states.

The covariance matrix of quantum fluctuation operators, in the thermodynamic limit, is defined as 
$$
G_{\mu\nu}(t)=\lim_{N\to\infty}G_{\mu\nu}^N(t)\, ,\qquad \mbox{where}\qquad  G_{\mu\nu}^N(t)=\frac{1}{2}\langle \left\{F_\mu^N,F_\nu^N\right\}\rangle_t \, .
$$
Another quantity of interest for such operators is the so-called symplectic matrix which encodes commutation relations between fluctuation operators. This is given as 
$$
s_{\mu\nu}(t)=\lim_{N\to\infty}s_{\mu\nu}^N(t)\, ,\qquad \mbox{where} \qquad s_{\mu\nu}^N(t)=-i\langle \left[F_\mu^N,F_\nu^N\right]\rangle_t\, .
$$
Due to the commutation relation between collective spin operators, the above matrix $s_{\mu\nu}^N$ is in fact a matrix which contains the expectation values of average operators. In particular, one has $s_{\mu\nu}^N=\sqrt{2}\sum_\eta \epsilon_{\mu\nu\eta}\langle m_\eta^N\rangle_t $. As such, the evolution of this matrix can be controlled through the mean-field equations for the average operators. This is not the case for the matrix $G(t)$, whose evolution is derived in what follows.  

It is convenient to introduce the matrix $K(t)$, whose entries  are given by 
\begin{equation}
K_{\mu\nu}(t)=\lim_{N\to\infty}K_{\mu\nu}^N(t)\, ,\qquad \mbox{where}\qquad K_{\mu\nu}^N(t)=\langle F_\mu^N F_\nu^N\rangle_t \, .
\end{equation}
The evolution of the covariance matrix can be recovered from the matrix $K(t)$ as $G(t)=[K(t)+K^T(t)]/2$ and we thus study for the moment the evolution of $K(t)$. We start from computing its time derivative. We have that 
$$
\dot{K}(t)=\lim_{N\to\infty}\frac{d}{dt}K_{\mu\nu}^N(t)=\lim_{N\to\infty}\left\langle \mathcal{L}^*\left[F_\mu^N F_\nu^N\right]\right\rangle_t \, ,
$$
where we exploited that $d/dt (F_\mu^N)=-\sqrt{N}\frac{d}{dt}(\langle m_\mu^N\rangle_t)$ is a scalar quantity and that $\langle F_\mu^N\rangle_t=0$. The task is thus to find an expression for the argument of the right-hand-side of the above equation. We note that
\begin{eqnarray}
\left\langle \mathcal{L}^*\left[F_\mu^N F_\nu^N\right]\right\rangle_t&=&i\left\langle [H,F_\mu^N ]F_\nu^N\right\rangle_t+i\left\langle F_\mu^N [H,F_\nu^N]\right\rangle_t+
\nonumber \\
&+&\sum_{\eta,\zeta}\frac{A_{\eta\zeta}}{2N}\left\langle \left[V_\zeta,\left[V_\eta,F_\mu^NF_\nu^N\right]\right]\right\rangle_t+
\nonumber\\
&+& i\sum_{\eta,\zeta}\frac{B_{\eta\zeta}}{2N}\left\langle \left\{V_\zeta,\left[V_\eta,F_\mu^NF_\nu^N\right]\right\}\right\rangle_t . 
\label{sm-fl:term}
\end{eqnarray}
We now derive the thermodynamic limit for all the different expectations appearing in the above equation. 

For the Hamiltonian term, we observe that 
$$
\left[H,F_\mu^N\right] =i\Omega \sum_\eta \epsilon_{x\mu\eta} \frac{V_\eta}{\sqrt{N}}\, ,
$$
and since $\langle F_\nu^N\rangle_t=0$ we can write, subtracting the expectation value of the above quantity, 
$$
i\left\langle [H,F_\mu^N]F_\nu^N\right\rangle_t =-\Omega \sum_\eta \frac{\epsilon_{x\mu\eta}}{\sqrt{N}} \left\langle V_\eta F_\nu^N\right\rangle_t =-\Omega \sum_\eta \epsilon_{x\mu\eta} \left\langle F_\eta^N F_\nu^N\right\rangle_t \, .
$$ 
Taking the thermodynamic limit, we find that 
$$
\lim_{N\to\infty} i\left\langle [H,F_\mu^N]F_\nu^N\right\rangle_t =\sum_\eta D_{\mu\eta} K_{\eta \nu}(t)\, ,\qquad D_{\mu\eta}=-\Omega \epsilon_{x\mu\eta} \, .
$$
Similarly, for the second term we find 
$$
i\left\langle F_\mu^N [H,F_\nu^N]\right\rangle_t =-\Omega \sum_\eta \epsilon_{x\nu\eta}\langle F_\mu^N F_\eta^N\rangle_t \, ,
$$
which in the thermodynamic limit gives 
$$
\lim_{N\to\infty}i\left\langle F_\mu^N [H,F_\nu^N]\right\rangle_t=\sum_\eta K_{\mu\eta}(t) D^T_{\eta\nu}\, .
$$

We now consider the third term in Eq.~\eqref{sm-fl:term}. We focus on 
$$
\frac{1}{N}\left\langle \left[V_\zeta,\left[ V_\eta,F_\mu^N F_\nu^N\right]\right]\right\rangle_t =\left\langle \left[F_\zeta^N,\left[ F_\eta^N,F_\mu^N F_\nu^N\right]\right]\right\rangle_t \, .
$$
Since we have written this in terms of commutators of fluctuation operators and since overall this scales as the product of average operators, we can calculate this term in the termodynamic limit using the symplectic matrix $s(t)$. Specifically, we find 
$$
\lim_{N\to\infty}\frac{1}{N}\left\langle \left[V_\zeta,\left[ V_\eta,F_\mu^N F_\nu^N\right]\right]\right\rangle_t=-s_{\zeta\mu}(t)s_{\eta \nu}(t)-s_{\zeta \nu}(t)s_{\eta\mu}(t)\, ,
$$
through which we can calculate
$$
\lim_{N\to\infty}\sum_{\eta,\zeta}\frac{A_{\eta\zeta}}{2N}\left\langle \left[V_\zeta,\left[V_\eta,F_\mu^NF_\nu^N\right]\right]\right\rangle_t=-[s(t)As(t)]_{\mu\nu}\, .
$$

We are then left with the last term on the right-hand-side of Eq.~\eqref{sm-fl:term}, which we divide into two pieces as 
\begin{equation}
\frac{i}{2N}\left\langle \left\{V_\zeta,\left[V_\eta,F_\mu^NF_\nu^N\right]\right\}\right\rangle_t=\frac{i}{2}\left\langle \left\{F_\zeta^N,\left[F_\eta^N,F_\mu^N F_\nu^N\right]\right\}\right\rangle_t +i\langle m_\zeta^N\rangle_t \langle \left[V_\eta,F_\mu^N F_\nu^N\right]\rangle_t \, .
\label{sm-fl:term-B}
\end{equation}
The second term emerges from the subtraction of the expectation values to the term $V_\zeta$ in the anticommutator. The first term on the right hand side above can be calculated using the commutation relation between fluctuation operators and the definition of the covariance matrix. Putting this back into the summation in Eq.~\eqref{sm-fl:term}, we find 
$$
\lim_{N\to\infty} \sum_{\eta,\zeta}\frac{iB_{\eta\zeta}}{2} \left\langle \left\{F_\zeta^N,\left[F_\eta^N,F_\mu^N F_\nu^N\right]\right\}\right\rangle_t=[s(t)BG(t)+G(t)Bs(t)]_{\mu\nu}
$$
The second term in Eq.~\eqref{sm-fl:term-B} looks like a Hamiltonian contribution and can indeed be treated in the same way in which we treated the terms involving $H$. By doing so, we find 
\begin{equation}
\lim_{N\to\infty}i\sum_{\eta,\zeta}{B_{\eta\zeta}}\langle m_{\zeta^N}\rangle_t\left\langle \left[V_\eta, F_\mu^N F_\nu^N\right]\right\rangle_t =[\bar{D}(t) K(t)]_{\mu\nu}-[K(t)\bar{D}(t) ]_{\mu\nu}\, ,
\end{equation}
where we have 
$$
\bar{D}_{\mu\nu}(t)=-\sum_{\eta,\zeta}\sqrt{2}B_{\eta\zeta} m_\zeta(t) \epsilon_{\eta\mu\nu}\, .
$$
Summing up all the contribution, we found that 
$$
\dot{K}(t)=[D(t)+\bar{D}(t)]K(t)-K(t)[D(t)+\bar{D}(t)]+s(t)BG(t)+G(t)Bs(t) -s(t)As(t)\, , 
$$
which for the covariance matrix becomes 
$$
\dot{G}(t)=W(t)G(t)+G(t)W^T(t)-s(t)As(t)\, , \qquad \mbox{with} \qquad W(t)=D(t)+\bar{D}(t)+s(t)B\, .
$$

\subsection{Products of collective operators}
Here, we show how to express the expectation values of products of collective operators of the type $\langle V_\mu V_\nu\rangle_t/N$ in terms of average operators (providing the leading-order extensive behavior) and of quantum fluctuations (providing the intensive correction to the leading order). 

We start from the expectation value of interest and we add and subtract the expectation, written in terms of average operators, to one of the collective operators as
\begin{equation}
\frac{1}{N}\langle V_\mu V_\nu\rangle_t =\frac{1}{N}\left\langle (V_\mu -N\langle m_\mu^N\rangle_t)V_\nu \right\rangle_t + \langle m_\mu^N \rangle_t \langle V_\nu\rangle_t\, .
\end{equation} 

Next we note that $V_\nu =Nm_\nu^N$ and thus we can write 
$$
\frac{1}{N}\langle V_\mu V_\nu\rangle_t =N\langle m_\mu^N \rangle_t \langle m_\nu^N\rangle_t+\frac{1}{N}\left\langle (V_\mu -N\langle m_\mu^N\rangle_t)V_\nu \right\rangle_t 
$$
Now we proceed by adding and subtracting $N$ times $\langle m_\nu^N\rangle$ to the operator $V_\nu$ in the second term on the right hand side of the above equation. We end up with (reordering the terms) 
$$
\frac{1}{N}\langle V_\mu V_\nu\rangle=N\langle m_\mu^N\rangle \langle  m_\nu^N\rangle+\frac{1}{N}\left\langle \left(V_\mu -N\langle m_\mu^N\rangle_t\right)\left(V_\nu-N\langle m_\nu^N\rangle_t\right)\right\rangle_t \, .
$$
Recalling the definition of the fluctuation operators $F_\mu^N$ we can write 
$$
\frac{1}{N}\langle V_\mu V_\nu\rangle =N\langle m_\mu^N\rangle_t \langle  m_\nu^N\rangle_t  +\langle F_\mu^N F_\nu^N \rangle_t\, .
$$
We conclude by noticing the algebraic relation 
$$
F_\mu^N F_\nu^N =\frac{1}{2}(F_\mu^N F_\nu^N+F_\nu^N F_\mu^N)+\frac{1}{2}[F_\mu^N,F_\nu^N]\, ,
$$
which taking the expectation  gives
$$
\langle F^N_\mu F^N_\nu\rangle_t =G^N_{\mu\nu}(t) +\frac{i}{2}s_{\mu\nu}^N(t)\, ,
$$
where $s^N_{\mu\nu}(t)$ is the (degenerate) symplectic matrix 
$$
s^N_{\mu\nu}(t)=-i\langle [F_\alpha,F_\beta]\rangle_t=\sqrt{2}\epsilon_{\mu\nu\eta } \langle m_\eta^N\rangle_t\, ,
$$
solely function of the average operators. 
Summarizing, we have found 
\begin{equation}
\frac{1}{N}\langle V_\mu V_\nu\rangle =N\langle m_\mu^N \rangle_t \langle m^N_\nu\rangle_t  +G^N_{\mu\nu}(t)+\frac{i}{2}s^N_{\mu\nu}(t)\, ,
\label{final}
\end{equation}
which allows us to express all thermodynamic quantities in terms of average operators and matrices associated with quantum fluctuation operators.

\subsection{von Neumann entropy from quantum fluctuation operators}
We now briefly discuss how to calculate the von Neumann entropy of the boundary time-crystal from the analysis of quantum fluctuations. This is basically achieved by considering that the latter behave as bosonic operators. 

The idea is as follows. The symplectic (degenerate) matrix $s(t)=\lim_{N\to\infty}s^N(t)$ can be brought, at any time $t$, into its canonical form via a rotation $R$ such that 
\begin{equation*}
\tilde{s}(t)=Rs(t)R^T=
%\begin{pmatrix}
\left(
\begin{array}{ccc}
    0 &1&0\\
    -1&0&0\\
    0&0&0
\end{array}
\right)
%\end{pmatrix}\, .
\end{equation*}
Through such a rotation, we can also define the matrix $\tilde{G}(t)=RG(t)R^T$. The rotation $R$ amounts to aligning the  $z$-axis of the reference frame with the principal direction identified by the mean-field state of the system. Since we consider the sector with maximal angular momentum, the matrix $\tilde{G}(t)$ has only the principal $2\times2$ minor that can be different from zero. Thus, the rotation $R$ provides the canonical mode of the bosonic quantum fluctuation system. Now, we can proceed as is done for Gaussian bosonic systems. We can define the matrix $\tilde{M}(t)=i\tilde{s}(t)\tilde{G}(t)$ and diagonalize it. As mentioned in the main text, the eigenvalues are given by $0,\pm\lambda$, with $\lambda>0$. The von Neumann entropy is then given by 
$$
S=\left(\lambda+\frac{1}{2}\right)\log \left(\lambda+\frac{1}{2}\right)-\left(\lambda-\frac{1}{2}\right)\log \left(\lambda-\frac{1}{2}\right)\, .
$$
For the sake of the calculation of the entropy, it is not necessary to find the rotation $R$ since the spectrum of $\tilde{M}(t)$ is equivalent to the spectrum of the matrix $M(t)=is(t)G(t)$. We note that for the thermal state which is obtained when $\Omega=0$, we have that $\tilde{G}={\rm diag}[n_\beta+1/2,n_\beta+1/2,0]$ and the entropy is thus given by 
$$
S_\beta=(n_\beta+1)\log(n_\beta+1)-n_\beta\log n_\beta\, .
$$

\section{Stochastic entropy production and the fluctuation theorem}
\label{App2}

In this Section we report the definition of the stochastic entropy production and the details of the associated quantum Crooks fluctuation theorem (FT). To this end we follow Ref.~\cite{KwonPRX2019}, see also Refs.~\cite{ManzanoPRX2018,LandiRMP2021,DeChiaraPRR2022} and references therein. Suppose a quantum system, initially in the state $\rho$, undergoes an evolution described by a quantum channel, i.e., a completely positive trace-preserving map, $\cal N$ such that its evolved state is ${\cal N}(\rho)=\sum_m K_m \rho K_m^\dagger$, where we have defined the Kraus operators $K_m$ satisfying the normalisation condition $\sum_m K_m^\dagger K_m =  \mathbb{1}$. Notice that the evolution generated by a Lindblad master equation, even a local one as Eq.~\eqref{Lindblad}, can be represented in this way. We also define a reverse map for the channel $\cal N$, following Crooks' prescription \cite{CrooksPRA2008}. To this end we first identify the fixed point $\pi$ for the channel $\cal N$: ${\cal N}(\pi)=\pi$. Then, we define the modified Kraus operators: $\tilde K_m = \pi^{1/2} K_m^\dagger \pi^{-1/2}$. The quantum reversed map is then defined as ${\cal R}(\rho)=\sum_m \tilde K_m \rho {\tilde K}_m^\dagger$. Notice that Ref.~\cite{KwonPRX2019} considers a more general quantum reversal map, called Petz's recovery map, defined in terms of an arbitrary reference state. For our purposes we find it sufficient to choose the reference state as the steady state $\pi$.

Let us assume for concreteness that the eigenvalue decomposition of the initial state $\rho$ reads:
\begin{equation}
    \rho=\sum_\mu p_\mu \ket{\psi_\mu}\bra{\psi_\mu}
\end{equation} 
in terms of its eigenvalues $p_\mu$ and eigenvectors $\psi_\mu$. Similarly for the evolved state:
\begin{equation}
    {\cal N}(\rho)=\sum_{\nu'} q_{\nu'} \ket{\phi_{\nu'}}\bra{\phi_{\nu'}}.
\end{equation} 
We also introduce the spectral decomposition for the steady state:
\begin{equation}
    \pi=\sum_i r_i \ket{i}\bra{i}.
\end{equation} 

When the system makes a transition from the initial eigenstate $\psi_\mu$ to the final eigenstate $\phi_{\nu'}$, the single-shot change of entropy is:
\begin{equation}
    \delta s^{\mu\to\nu'}=-\log q_{\nu'} + \log p_\mu.
\end{equation}
If the evolution were classical, this change of entropy would reduce to the usual change of the Shannon entropy of the eigenvalues of the initial and final density matrices. However, in the quantum case, the initial state $\rho$ may contain and develop quantum coherences in the eigenbasis $\{\ket i\}$ of the steady state $\pi$. Ref.~\cite{KwonPRX2019} thus defines the quantum information exchange:
\begin{equation}
\delta q_{ij\to kl} = -\log\left[\frac{r_{k}r_{l}}{r_i r_j} \right],    
\end{equation}
associated with the process $\ket i\bra j %\xrightarrow
\to_{\cal N} \ket{k}\bra{l}$. For a classical evolution of a system in contact with an equilibrium reservoir, where the evolution is restricted to diagonal states, the quantity $\delta q$ is related to the stochastic heat exchange for a particular transition. 

With these premises, we introduce the stochastic entropy production for the transition $(\mu,i,j)\to (\nu',k,l)$ as
\begin{equation}
    \sigma^{\mu\to\nu'}_{ij\to k' l'}=\delta s^{\mu\to\nu'} - \delta q_{ij\to kl}.
\end{equation}
The associate complex-valued quasi-probability for this value is:
\begin{equation}
P^{\mu,\nu'}_{ij,kl} = p_\mu \bra{\phi_{\nu'}} \Pi_{k} {\cal N}(\Pi_i \ket{\psi_\mu}\bra{\psi_\mu}\Pi_j)\Pi_{l}\ket{\phi_{\nu'}},
\end{equation}
where $\Pi_i=\ket i\bra i$ is a projector onto an eigenstate of the steady state $\pi$.

The probability distribution of the stochastic entropy production for the forward process can be then written as:
\begin{equation}
    P(\sigma) = \sum_{\mu,i,j}\sum_{\nu',k,l} P^{\mu,\nu'}_{ij,kl} \delta(\sigma-\sigma^{\mu\to\nu'}_{ij\to k' l'}),
\end{equation}
and similarly for the backward process $P_R(\sigma)$ using the reversal map $\cal R$.
Notice that, even though $P^{\mu,\nu'}_{ij,kl}$ may be complex, it is possible to show that $P(\sigma)$ and $P_R(\sigma)$ are real-valued but not necessarily positive, due to the quantum coherence of the states $\rho$ and $\cal N(\rho)$ in the eigenbasis of the reference state $\pi$. Therefore $P(\sigma)$ and $P_R(\sigma)$ should be regarded as quasi-probability distributions, fulfilling the quantum Crooks relation:
\begin{equation}
    \label{SM:Crooks}
    \frac{P(\sigma)}{P_R(-\sigma)}=e^{\sigma},
\end{equation}
as a detailed FT whose proof can be found in Ref.~\cite{KwonPRX2019}.
If we rearrange the terms in Eq.~\eqref{SM:Crooks} and integrate we obtain the corresponding integral FT:
\begin{equation}
    \label{SM:integralFT}
    \left\langle e^{-\sigma} \right \rangle = \int  e^{-\sigma} P(\sigma) d\sigma=1. 
\end{equation}

\begin{figure}[t]
\centering
\includegraphics[width=0.5\columnwidth]{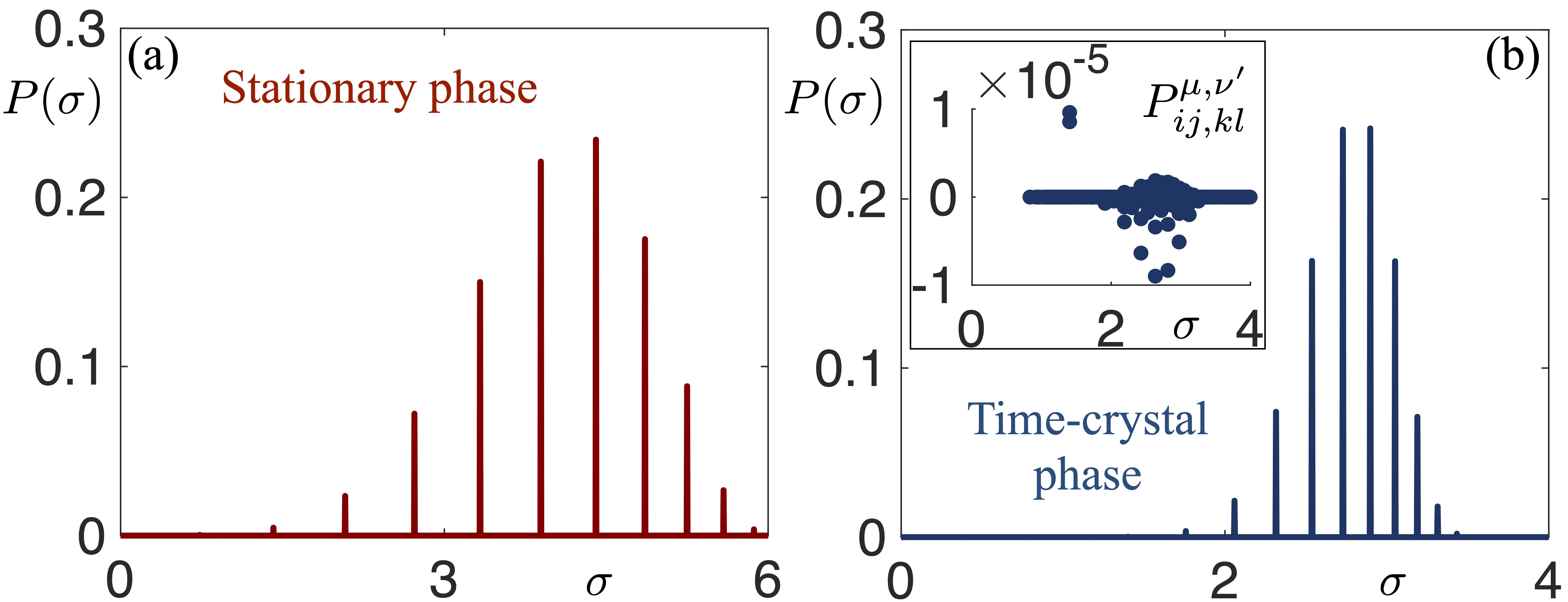}
\caption{{\bf Stochastic entropy production} (a) Quasi-probability $P(\sigma)$ for observing an entropy production $\sigma$ in the (a) stationary phase ($\Omega/\Gamma=0.67$) and in the (b) time-crystal regime ($\Omega/\Gamma=2$) for $N=10$ and $n_\beta=1$. The inset of (b) shows how negative quasi-probabilities $P_{ij,kl}^{\mu,\nu'}$ are possible in the time-crystal phase which are absent in the stationary phase. }
\label{Fig3}
\end{figure}

Fig.~\ref{Fig3} shows the quasi-probability distribution $P(\sigma)$ in the stationary and time crystal phases. The discrete nature of the distribution is typical of systems with a finite, albeit large, number of states. We also notice that the values of the stochastic variable $\sigma$ are always positive even though the second law would in principle allow for a small probability of negative $\sigma$. 
The inset of Fig.~\ref{Fig3}(b) also shows the occurrence of  negative quasi-probabilities $P_{ij,kl}^{\mu,\nu'}$ in the time-crystal phase. 
In contrast, in the stationary phase, negative quasi-probabilities seem not to occur (negative values of the order $10^{-10}$ can be observed but we associated them with numerical errors).

%%%%%%%%%%%%%%%%%%%%%%%%%%%%%%%%%%%%%%%%%%%%%%%%
%%%%%%%%%%%%%%%%%%%%%%%%%%%%%%%%%%%%%%%%%%%%%%%%

\section*{References}
\bibliographystyle{iopart-num}
\bibliography{refs}

\end{document}